\documentclass[aps,pra,twocolumn,tightenlines,showkeys,showpacs,groupedaddress]{revtex4}
\usepackage{graphicx}
\usepackage{amsmath}
\usepackage{longtable}

\begin{document}

\title{Convergence of an $s$-wave calculation of the He ground state.}
\author{J.Mitroy}
\email{jxm107@rsphysse.anu.edu.au}
\affiliation{Faculty of Technology, Charles Darwin University, Darwin NT 0909, Australia}
\author{M.W.J.Bromley}
\email{mbromley@physics.sdsu.edu}
\affiliation{Department of Physics, San Diego State University, San Diego CA 92182, USA}
\author{K.Ratnavelu}
\email{kuru052001@gmail.com}
\affiliation{Institute of Mathematical Sciences, University of Malaya, 50603 Kuala Lumpur, Malaysia}

\date{\today}

\begin{abstract}

The Configuration Interaction (CI) method using a large Laguerre 
basis restricted to $\ell = 0$ orbitals is applied to the
calculation of the He ground state.  The maximum number of orbitals 
included was 60.  The numerical evidence suggests that the energy 
converges as $\Delta E^N \approx A/N^{7/2} + B/N^{8/2} + \ldots$ 
where $N$ is the number of Laguerre basis functions.  The 
electron-electron $\delta$-function expectation converges as 
$\Delta \delta^N \approx A/N^{5/2} + B/N^{6/2} + \ldots$ 
and the variational limit for the $\ell = 0$ basis is estimated 
as $0.1557637174(2)$ $a_0^3$.  It was seen that extrapolation of 
the energy to the variational limit is dependent upon the basis 
dimension at which the exponent in the Laguerre basis was optimized. 
In effect, it may be best to choose a non-optimal exponent if one 
wishes to extrapolate to the variational limit. An investigation 
of the Natural Orbital asymptotics revealed the energy converged 
as $\Delta E^N \approx A/N^{6} + B/N^{7} + \ldots$ while the 
electron-electron $\delta$-function expectation converged as 
$\Delta \delta^N \approx A/N^{4} + B/N^{5} + \ldots$.  The
asymptotics of expectation values other than the energy showed 
fluctuations that depended on whether $N$ was even or odd.  

\end{abstract}


\maketitle

\section{Introduction}

There have been a number of studies of the convergence of the 
configuration interaction (CI) expansion of the helium ground state 
\cite{carroll79a,hill85a,kutzelnigg92a,decleva95a,jitrik97a,ottschofski97a,sims02a,bromley06a}  
following the pioneering work of Schwartz \cite{schwartz62a}.   
These studies have investigated the convergence of the energy with
respect to the number of partial waves included in the wave function 
and also with respect to the dimension of the radial basis.

It has been known since 1962 \cite{schwartz62a} that the energy 
converges slowly with respect to $J$, the maximum angular momentum 
of any orbital included in the CI expansion.  In particular the 
leading term to the energy increment is known to behave as 
\begin{equation} 
\Delta E^{J}  =  \langle E \rangle^J - \langle E \rangle^{J-1} \sim  \frac{A_E}{(J+{\scriptstyle \frac{1}{2}})^4}  \label{pE1} \\   
\end{equation} 
at high $J$.  Later work 
\cite{carroll79a,hill85a,kutzelnigg92a,decleva95a,ottschofski97a} 
showed that the energy increments can be written more generally as 
\begin{equation} 
\Delta E^J = \frac{A_E}{(J+{\scriptstyle \frac{1}{2}})^4} 
   + \frac{B_E}{(J+{\scriptstyle \frac{1}{2}})^{5}}
   + \frac{C_E}{(J+{\scriptstyle \frac{1}{2}})^{6}} + \ldots \ , 
\label{Eseries} 
\end{equation} 
where explicit expressions for $A_E$ and $B_E$ exist, namely  
\begin{eqnarray} 
A_E &=& -6\pi^2 \int |\Psi(r,r,0) |^2 r^5 dr = -0.074 226  \label{AE} \\  
B_E &=& -\frac{48 \pi}{5} \int |\Psi(r,r,0) |^2 r^6 dr = -0.030 989  \label{BE} \ .  
\end{eqnarray} 
No expressions for $C_E$ exist. The numerical values in eqs.~(\ref{AE}) 
and (\ref{BE}) are obtained from close to exact wave functions 
\cite{hill85a}. 

However, the convergence with respect to $J$ represents only one
aspect of the convergence problem.  Just as important is the 
convergence with respect to the dimension of the radial basis $N$,
for a given $J$. How do the increments to $E$ with increasing 
$N$ 
\begin{eqnarray} 
\Delta E^{N} & = & \langle E \rangle^N - \langle E \rangle^{N-1}    \nonumber \\  
           & \sim & \frac{A'_E}{N^p} + \frac{B'_E}{N^{p+t}} + \frac{C'_E}{N^{p+2t}} + \ldots \ , \label{pEN}    
\label{ENseries} 
\end{eqnarray} 
behave?  In effect, what are the values of $p$ and $t$?   This aspect 
of the CI expansion is not as well understood as the convergence with 
$J$ and  there have been no studies equivalent in sophistication to 
those of Schwartz \cite{schwartz62a}, Hill \cite{hill85a} and  
Kutzelnigg and collaborators \cite{kutzelnigg92a,ottschofski97a}.
Some attention has been given to the radial convergence of the 
hydrogen atom in gaussian basis sets \cite{kutzelnigg94a}.   
The seminal investigation of Carroll and collaborators concluded
that $p \approx 6$ for a natural orbital (NO) basis 
\cite{lowdin56a,carroll79a}.  This result has been quite influential, 
and can be regarded as ultimately motivating the use of principal 
quantum number expansions to 
extrapolate energies to the infinite basis limit from correlation 
consistent basis sets \cite{klopper99a}.  More recently, Goldman 
performed a regression analysis to give $p \approx 5.7$ for a 
NO basis and $p \approx 3.8$ for a Slater basis with a common 
exponent \cite{goldman95a}.   

\begingroup
\begin{table*}[bth]
\caption[]{ Comparison of different CI calculations of the 
$s$-wave model of the He atom ground state.   The expectation 
value of the electron-electron $\delta$-function (in $a_0^3$) 
is denoted as $\langle \delta \rangle$.  The data in the 
$\langle E \rangle^M$ and $\langle E \rangle^{60}$ columns are 
the energies (in hartree) with $N=M$ and $N=60$ basis sets 
respectively.  The data in the $\langle E^{\infty} \rangle$ 
and $\langle \delta \rangle^{\infty}$ columns are obtained by   
doing an explicit calculations with $N = 60$ and then adding
in the $60 \to \infty$ correction assuming an $A/N^{-p}$ 
asymptotic form.  
}
\label{Hetab1}
\vspace{0.2cm}
\begin{ruledtabular}
\begin{tabular}{lcccccc}
$M$ & $\lambda_{M}$  & $\langle E \rangle^M$ & $\langle E\rangle^{60}$ &  $\langle E \rangle^{\infty}$ & $\langle \delta \rangle^M$ &  $\langle \delta \rangle^{\infty}$ \\ \hline   
 10 & 3.07   &  -2.879 022 691 296 &  -2.879 028 727 964   & -2.879 028 7667  &  0.155 922 600 334 &  0.155 763 879  \\
 20 & 4.80   &  -2.879 028 507 141 &  -2.879 028 754 899   & -2.879 028 7671  &  0.155 789 345 524 &  0.155 763 796  \\
 30 & 6.45   &  -2.879 028 726 601 &  -2.879 028 761 447   & -2.879 028 7672  &  0.155 772 304 341 &  0.155 763 771  \\
 40 & 8.04   &  -2.879 028 756 467 &  -2.879 028 763 935   & -2.879 028 7673  &  0.155 767 637 040 &  0.155 763 760  \\
 50 & 9.57   &  -2.879 028 763 441 &  -2.879 028 765 118   & -2.879 028 7670  &  0.155 765 847 239 &  0.155 763 757  \\
 60 & 11.10  &  -2.879 028 765 650 &  -2.879 028 765 650   & -2.879 028 7661  &  0.155 765 002 122 &  0.155 763 860 \\
\multicolumn{3}{l}{Decleva {\em et al} \cite{decleva95a}}  &  -2.879 028 767 289 \\ 
\multicolumn{3}{l}{Goldman {\em et al} \cite{goldman95a}}  &  -2.879 028 767 319 \\ 
\end{tabular} 
\end{ruledtabular}
\end{table*}
\endgroup

The radial basis sets used for the configuration interaction or 
many body perturbation theory treatments of atomic structure 
can be broadly divided into two classes.  In the first
class, one defines a box and a piece-wise polynomial (e.g a spline) 
is used to define the radial dependence of the wave function in the
interior of the box.  The properties of the radial basis are
determined by the size of the box, the number of knot points, and 
where they are located.  The other approach typically expands the 
wave function in terms of a basis of functions with a convenient 
analytic form, examples would be an evenly tempered set of Slater 
type orbitals (STOs) \cite{kutzelnigg92a,sims02a} (this type of 
basis set is often optimized 
with respect to a couple of parameters used to defined a sequence 
of exponents) or a set of Laguerre type orbitals (LTOs) 
\cite{shull55a,holoien56a,bromley02a}.  

The two most recent examples of these two approaches are the calculations 
by Decleva {\em et al} (B-splines) \cite{decleva95a} and Sims and Hagstrom 
(Slater basis) \cite{sims02a} which are the biggest calculations
of their respective type.  The B-spline calculation has given estimates
$\Delta E^{J}$ increments that are believed to be accurate to within 
$10^{-8}$ hartree or better.  One of the reasons this accuracy is 
possible is that $\Delta E^{J}$ varies smoothly as the number of knot
points is adjusted.  This made it possible to obtain reasonable 
estimates of the infinite basis limit.  Their estimate of the $s$-wave
limit was accurate to better than 10$^{-9}$ hartree.   Achieving this 
extreme level of accuracy was not possible when using the Slater basis 
\cite{sims02a} since linear dependence issues made it problematic to 
expand their radial basis to completeness. Indeed, resorting to REAL*24
arithmetic still resulted in an error of $4 \times 10^{-6}$ hartree.  

In some investigations of the convergence properties of the CI expansion
for the helium atom \cite{bromley06a} and mixed electron-positron systems 
\cite{mitroy06a} it became apparent that a better understanding of the 
issues that influence the convergence of the radial basis was desirable.
For example, it was apparent that the dimension of the radial basis
should be increased  as $J$ increases in order to ensure the successive 
$\Delta E^{J}$ increments are computed to the same relative accuracy 
\cite{kutzelnigg99a,mitroy06a,bromley06a}.  In addition, it was readily 
apparent that extrapolation of the radial LTO basis to the $N \to \infty$ 
limit was not straightforward.  

In this work, we investigate the radial convergence of the CI expansion 
for a more manageable model of the helium atom with the orbitals restricted 
to the $\ell = 0$ partial wave.   The linear dependence issues that are 
such a problem for a Slater basis are eliminated by choosing the radial 
basis to consist of LTOs \cite{goldman89a,bromley02a}, (formally, the 
LTO basis spans the same space as the common exponent Slater basis, 
i.e. $r^{n_i} \exp(-\lambda r)$).  We note in passing the previous
work of Hol{\o}ein \cite{holoien56a} who also investigated the convergence
of a (small) LTO basis for an $s$-wave model of helium.   Initially, 
we examine the merits of using an LTO basis with the exponent optimized 
to the basis dimension.  The nature of the asymptotic expansion for 
the energy increments is then deduced.  Finally, the density matrix for our
best wave function is diagonalized and the convergence properties of 
the natural orbital expansion are also determined.  As part of this
analysis, attention is also given to the convergence of the
electron-electron coalescence matrix element since it arises in 
calculations of the two-electron relativistic Darwin correction 
\cite{halkier00a} and electron-positron annihilation \cite{mitroy02b}.

\section{The $s$-wave energy for helium}

The non-relativistic hamiltonian for the $^1S^e$ ground state of 
helium 
\begin{equation} 
H  =  - \sum_{i=1}^{N_e} \left( \frac {1}{2} \nabla_{i}^2 + \frac{2}{r_i} \right) 
+ \frac{1}{r_{12}} \ , 
\end{equation} 
is diagonalized in a basis consisting of anti-symmetric products of 
single electron orbitals 
\begin{eqnarray}
|\Psi;S=0 \rangle = \sum_{i,j} c_{ij} \: \mathcal{A}_{12} \:
     \langle {\scriptstyle \frac12} \mu_i {\scriptstyle \frac12} \mu_j|0 0 \rangle                                                                                                  
   \phi_i({\bf r}_1) \phi_j({\bf r}_2)  .
\label{wvfn}
\end{eqnarray}
The functions $\phi({\bf r})$ are single electron orbitals
Laguerre functions with the radial form  
\begin{equation}
\chi _\alpha (r)=N_\alpha r^{\ell} \exp (-\lambda _\alpha r)
L_{n_\alpha -\ell - 1}^{(2\ell +2)}(2\lambda _\alpha r) \ ,
\label{LTO}
\end{equation}
where the normalization constant is
\begin{equation}
N_\alpha =\sqrt{\frac{(2\lambda_\alpha) (n_\alpha-\ell-1)!}
{(\ell+n_\alpha+1)!}} \ .
\label{LTOnorm}
\end{equation}
The function $L_{n_\alpha-\ell-1}^{(2\ell +2)}(2 \lambda _\alpha r)$ is 
an associated Laguerre polynomial that can be defined in terms of a 
confluent hypergeometric function \cite{abramowitz72a,bromley02a,bromley02b}.
In the present work $\ell$ is set to zero.  The basis can be characterized 
by two parameters, $N$, the number of LTOs, and $\lambda$ the exponent 
characterizing the range of the LTOs.  It is normal to use a common 
exponent, $\lambda$ and when this is done the basis functions form 
an orthogonal set.  The exponent can be optimized to give the lowest
energy and when this is done  one can define $\lambda_M$ 
to be the value of optimal $\lambda$ for a basis
of dimension $M$.  The value of $\lambda_M$ was seen to increase $M$ 
increased.      

Calculations for basis sets with $\lambda_M$ optimized for dimensions 
of 10, 20, 30, 40, 50 and 60 have been performed.  The optimal exponents 
are listed in Table \ref{Hetab1} along with the energy for those
values of $M$.   The calculations were expedited by using the result that
the two-electron Slater integrals for any $\lambda$ are related to 
each other by a very simple scaling relation. Accordingly, the list of
Slater integrals could be generated once (by numerical integration
using gaussian quadrature \cite{bromley02a}) for a given $\lambda$, and
then recycled by rescaling for calculations at different $\lambda$
(refer to Appendix B).
Calculations with $N$ ranging from 1 to 60 have been 
performed for the six values of $\lambda_M$ listed in Table \ref{Hetab1}.  
The dimension of the hamiltonian for the largest calculation was 1830.  
The quantities listed in the tables and the text are given in atomic units.
The most precise energy for the helium $s$-wave model is that of Goldman
\cite{goldman94a,goldman95a} who used a basis written in terms of
$r_{<}$, $r_{>}$ co-ordinates to obtain an energy of 
$E = -2.879 028 767 319 214$ hartree.
 
\subsection{Use of quadruple precision arithmetic}

The present calculations were all performed with quadruple
precision arithmetic.  It was only possible to get energies
precise to 13 significant digits for the largest calculations 
when double precision arithmetic was used.  This was caused 
by roundoff error gradually accumulating during the course
of the rather extensive calculations and the 13 digits appears 
to be the limit that can be achieved for double precision 
arithmetic (some experimentation revealed that the last 2 
digits of the 15 double precision digits were sensitive to 
different Fortran compilers and even the optimization options 
of those compilers).  The analysis requires investigation of 
the energy differences of eq.~(\ref{ENseries}), and these
energy differences can be rather small (e.g. 
$\Delta E^{60} = 1.5 \times 10^{-10}$ hartree for the 
$\lambda_{60}$ basis).  The fluctuations caused by roundoff
did have a noticeable impact on the parameters derived from
these energy differences at large $N$. These fluctuations
were removed once quadruple precision arithmetic was adopted.   

\section{Simple power law decay}

All observable quantities can be defined symbolically as  
\begin{equation}
\langle X \rangle^{N} = \sum_{n=1}^{N} \Delta X^{n} \ ,  
\label{XN1}
\end{equation}
where $\Delta X^{n}$ is the increment to the observable that occurs
when the basis dimension is increased from 
$n - 1$ to $n$, e.g.     
\begin{equation}
\Delta X^{n} = \langle X \rangle^{n} - \langle X \rangle^{n-1} \ . 
\label{XN2}
\end{equation}
Hence, one can express the limiting value formally as  
\begin{equation}
\langle X \rangle = \langle X \rangle^{N}  + \sum_{n=N+1}^{\infty} \Delta X^{n} \ .  
\label{XN3}
\end{equation}
The first term on the right hand side will be determined by explicit 
computation while the second term will be estimated.  Obtaining an 
estimate of the remainder term does require some qualitative knowledge of
how the $\Delta X^{N}$ terms decay with $N$.  For example,  previous 
computational investigations indicate that the natural orbital decomposition 
leads to a $\Delta E^{N} \approx N^{-6}$ dependence \cite{bunge70a,carroll79a}.  
As far as we know, there have not been any detailed investigations of the 
$N$ dependence of a Laguerre basis.  

A useful way to analyze the convergence is to assume the increments 
obey a power law decay of the form
\begin{equation}
\Delta X^{N}  \sim  \frac{A_X}{N^p} \ ,      
\label{Xpdef} 
\end{equation}
and then determine the value of $p$ from two successive 
values of $\Delta X$ using 
\begin{equation}
p =   \ln \left(  \frac {\Delta X^{N-1}}{\Delta X^N} \right) \biggl/   
      \ln \left( \frac{N}{N-1} \right) \ .  
\label{pdef} 
\end{equation}

\begin{figure}[bth] 
\centering
\includegraphics[width=8.5cm,angle=0]{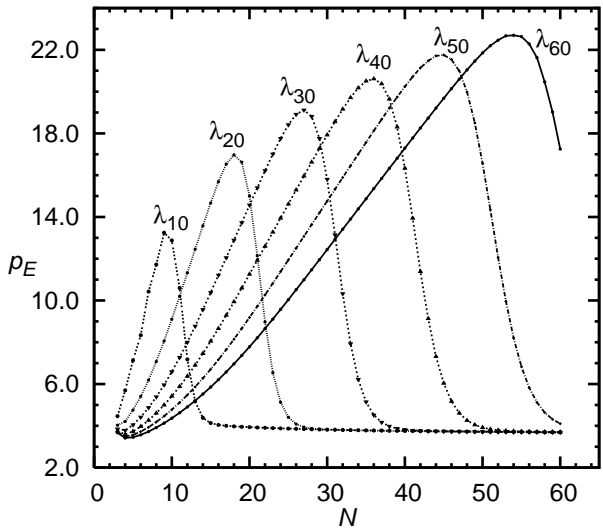}
\caption[]{
The exponents $p_E$ as a function of $N$ for the $s$-wave 
calculations of the He ground state.  The different curves
were obtained from the LTO basis sets with the exponents listed 
in Table \ref{Hetab1}. 
}
\label{pEHe}
\end{figure}

Figure \ref{pEHe} plots the exponent derived from the energy increments
for six different values of $\lambda_M$.   The succession of
curves show that $p_E$ tends to peak at values larger than 10 at an
intermediate $N$ and then shows a tendency to decrease.  The value at
$N = 60$ was $p_E \approx 3.7$ for the most of the curves shown in figure
\ref{pEHe}.

The salient point to be extracted from Figure \ref{pEHe} is that 
the value of $p_E$ for a given $\lambda_{M}$ at $N = M$ is quite 
different from the asymptotic value, e.g. the value of $p_E$ for 
the $\lambda_{20}$ curve is much larger at $N = 20$ than it is 
at $N = 60$. This is quite an 
annoying result.  Ideally, one would like to perform the largest 
calculation with the exponent optimized for that dimension basis.  
Then the specific form of the power law decay would be estimated 
by analyzing the energies obtained from a series of slightly 
smaller calculations.  This information would subsequently be used to 
estimate the energy or other expectation value in the variational
limit.  However, this is not possible since
the energy increments will not have achieved their asymptotic 
form.

\begin{figure}[th] 
\centering
\includegraphics[width=8.5cm,angle=0]{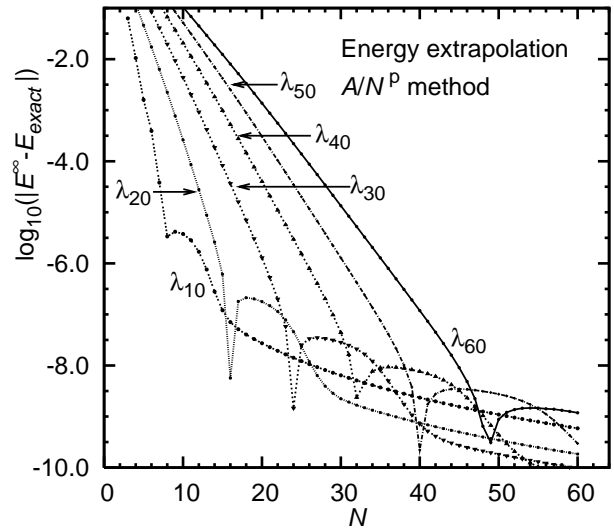}
\caption[]{
The extrapolated $N \rightarrow \infty$ limit for the He ground 
state energy for different values of $\lambda_M$.   
The exact $s$-wave energy is taken from the $J = 0$ calculation 
of Goldman \cite{goldman95a}. 
}
\label{HeEinf}
\end{figure}

Although there are problems in using an optimized exponent, it 
may still be possible to analyze a sequence of energies from a 
calculation with a non-optimized exponent and thereby estimate
the variational limit.  Assuming that the increments obey 
eq.~(\ref{Xpdef}), one can write  
\begin{equation}
A_X =   N^p \ \Delta X^{N} \ , 
\label{AEdef} 
\end{equation}
and thus the $n>N$ remainder term 
\begin{equation}
\sum_{n=N+1}^{\infty} \frac{A_X}{N^p } \approx  \frac{A_X}{(p-1)(N+{\scriptstyle \frac{1}{2}})^{p-1} }  \ .
\label{remainder}
\end{equation}
can be derived from $\langle X \rangle^{N-2}$, $\langle X \rangle^{N-1}$ 
and $\langle X \rangle^{N}$ \cite{bromley06a,mitroy06a}.    
When this remainder was evaluated in this work, the first 10000 terms 
of the sum over $n$ were computed explicitly.  Then the approximate 
relation eq.~(\ref{remainder}) was used. 

Figure \ref{HeEinf} shows the estimated variational limit as a function
of $N$ for the $\lambda_i$ listed in Table \ref{Hetab1}.  An explicit 
calculation including $N$ LTOs was initially performed to determine 
$\langle E \rangle^{N}$. Then eq.~(\ref{remainder}) was used to estimate
the remainder and hence deduce the variational limit.  The variational
limits in Table \ref{Hetab1} were extracted from the calculations with
$N = 60$.   The exact variational limit can be predicted to the 9th digit 
after the decimal point.  The most inaccurate estimate of the 
variational limit is that from the $\lambda_{60}$ calculation.   
So the calculation that is explicitly optimized at $N = 60$, 
(i.e. with $\lambda_{60}$), and gives the best energy at $N = 60$, 
gives the worst estimate of the variational limit!    

A CI calculation of the Li$^+$ ground state restricted to the
$\ell = 0$ partial wave was also performed to check whether the 
conclusions above were peculiar to He.  Once again, the exponent 
$p_E$ was approximately 3.7 at $N = 60$ and the convergence pattern 
for $N < M$ was distorted by optimization of the exponent. 

\begin{figure}[th] 
\centering
\includegraphics[width=8.5cm,angle=0]{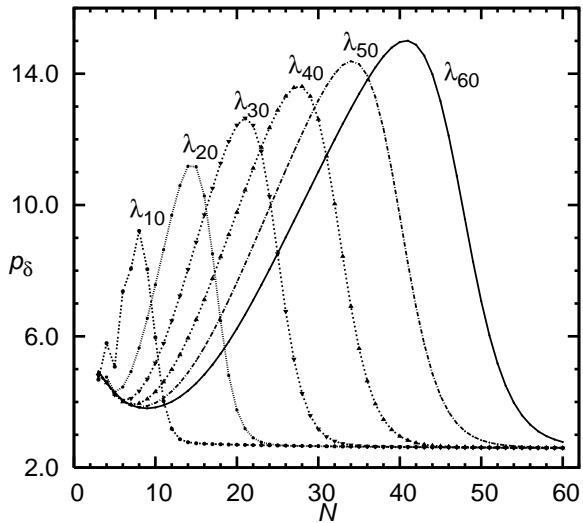}
\vspace{0.1cm}
\caption[]{
The estimated exponents $p_{\delta}$ as a function of $N$ for the 
LTO calculations of the He ground state $\langle \delta \rangle$.
The different curves were obtained with the LTO basis sets  
listed in Table \ref{Hetab1}. 
}
\label{pdHe}
\end{figure}

\section{The $\delta$-function expectation value} 

Part of the motivation for the present study is to gain a better
understanding of how to perform CI calculations for mixed 
electron-positron systems.   Apart from the energy, the next most 
important expectation value for a positronic system is the 
electron-positron annihilation rate \cite{mitroy02b}.  The 
annihilation rate is 
proportional to the expectation of the electron-positron delta 
function, and has the inconvenient property that it is even more 
slowly convergent than the energy \cite{bromley02a,bromley02e}.  
Accordingly, the convergence of the electron-electron 
$\delta$-function is investigated using the methodology 
previously used for the energy.   The only independent investigation 
of this quantity for an $s$-wave model of helium was by Halkier  
{\em et al} \cite{halkier00a} who obtained 0.155786 $a_0^3$.   

\begin{figure}[th] 
\centering
\includegraphics[width=8.5cm,angle=0]{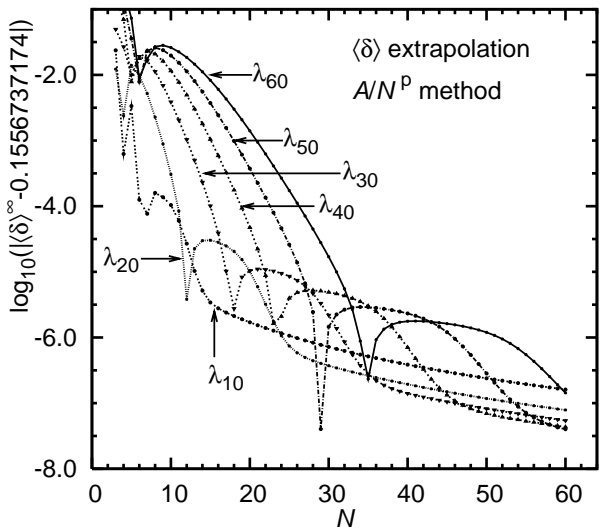}
\vspace{0.1cm}
\caption[]{
The convergence of $\langle \delta \rangle$ for the He ground state 
as a function of $N$.  The absolute value of the difference of the 
extrapolated $N \rightarrow \infty$ limit subtracted from $0.1557637174$ 
$a_0^3$ is plotted.       
}
\label{Hedinf}
\end{figure}

Figure \ref{pdHe} tracks the behavior of the exponent $p_{\delta}$ 
derived from eq.~(\ref{pdef}). It can be seen that $p_{\delta}$ achieves 
values exceeding 10 before it decreases to its asymptotic values.  
The present calculations give $p_{\delta} \approx 2.6$ at $N=60$ 
although $p_{\delta}$ is still exhibiting a slow but steady decrease.  

Although distortions in the convergence pattern are still present,
they are less severe than the energy since the successive 
$\Delta \delta^N$ increments are larger.  As a 
rule, $p_{\delta}$ was at least 10$\%$ larger than 2.6 at $N = M$.
A choice of $N \ge (M\!+\!10)$ would generally lead to $p_{\delta}$ 
being in the asymptotic region. 

Figure \ref{Hedinf} shows the estimated variational limit of 
$\langle \delta \rangle^{\infty}$ as a function of $N$ for the six 
different values of $\lambda_M$ listed in Table  \ref{Hetab1}.   
An explicit calculation including $N$ LTOs was initially performed, 
then eq.~(\ref{remainder}) was used to estimate the variational limit.   
A variational limit of $\langle \delta \rangle = 0.1557637174(2)$ $a_0^3$ 
(see later discussion) was assumed for plotting purposes.  The notable
feature is that the $\lambda_{60}$ estimate of the limit at $N = 60$ 
is one of the least accurate. 

\section{A closer look at the asymptotic power laws} 
  
\begin{figure}[th] 
\centering
\includegraphics[width=8.5cm,angle=0]{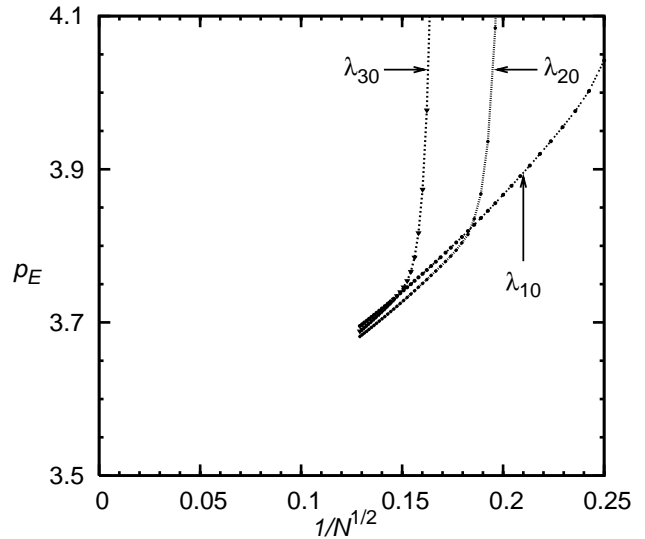}
\caption[]{
The exponents $p_E$ as a function of $\frac{1}{\sqrt{N}}$ for the $s$-wave 
calculations of the He ground state.  The different curves were obtained 
from the LTO basis sets with the exponents listed 
in Table \ref{Hetab1}. 
}
\label{pEHe2}
\end{figure}

\begin{figure}[th] 
\centering
\includegraphics[width=8.5cm,angle=0]{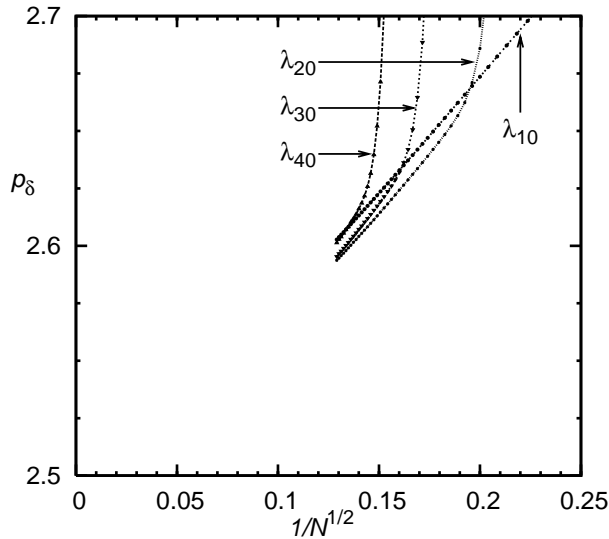}
\vspace{0.1cm}
\caption[]{
The estimated exponents $p_{\delta}$ as a function of $\frac{1}{\sqrt{N}}$ 
for the LTO calculations of the He ground state $\langle \delta \rangle$.
The different curves
were obtained from the LTO basis sets with the exponents listed 
in Table \ref{Hetab1}. 
}
\label{pdHe2}
\end{figure}

Figures \ref{pEHe2} and \ref{pdHe2} show the behavior of the $p_E$ and 
$p_\delta$ versus $\frac{1}{\sqrt{N}}$ for values of $N$ greater than 
16.  Curves are not shown for all the $\lambda_M$ exponents.  In some
cases, the values of $p_\delta$ did not fall within the plotting window.  
   
The notable feature to be gleaned from both set of curves is the essentially 
linear behavior $p_{\rm delta}$ with respect to $\frac{1}{\sqrt{N}}$ for 
values of $N$ greater than 16 and a visual inspection suggests that the 
limiting exponents is $p_{\delta} = 2.5$.  The purely visual evidence
that the lowest order term of $p_E$ is $O(N^{-1/2})$ is not as compelling 
as that for $p_{\rm delta}$, but by analogy with this form has been
assumed and subjected to extensive testing.

More substantial evidence is
provided by a fit of the $p$ vs $N$ data to an inverse power series 
of the form      
\begin{eqnarray} 
p &=& p_0 + \sum_{i=1}^{N_p} \frac{p_i}{\sqrt{N^{i}}} \ .   
\label{panalysis} 
\end{eqnarray} 
What we have done is fit $(N_p+1)$ successive $p_E$ or $p_{\delta}$ 
values to eq.~(\ref{panalysis}) for the $\lambda_{10}$ data sequence.
The results of those fits are given in Table \ref{Nlimits}. 

Using a 4 or 5 term fits (and 6 term fit in the case of $p_E$) 
results in limiting exponents very close to either 3.5 
(for $p_E$) or 2.5 (for $p_{\delta}$).    
These estimates were also reasonably stable.  For example,
the value of $p_E$ for a 5-term fit for a data sequence 
terminating at $N = 50$ (as opposed to $N = 60$ in Table  
\ref{Nlimits}) was 3.4970.   

\begingroup
\begin{table}[bth]
\caption[]{ Results of using 2, 3, 4, 5 or 6 term inverse power series
(corresponding to $N_p = 1, 2, 3, 4$ and 6) 
to determine the limiting value of the exponents and the energy
and $\delta$-function (in $a_0^3$).  The results for the LTO
data sequences were taken from the largest calculations for the 
$\lambda_{10}$ exponents.  The results for the NO data sequences 
were extracted at $N = 20$ while those for the energy optimized LTO
sequence were determined at $N = 30$. Data entries with an asterisk, 
$^*$, were obtained using a weighted average (as described in the text) 
due to fluctuations depending on whether the $N$ was even or odd.  
}
\label{Nlimits}
\vspace{0.2cm}
\begin{ruledtabular}
\begin{tabular}{lcccc}
$N_p$ & $p_0$  & $p_1$  & $A_X$ &  $\langle X \rangle^{\infty}$  \\ \hline   
    \multicolumn{5}{c}{LTO data sequence: $\lambda_{10}$ basis} \\  
    \multicolumn{5}{c}{$\langle E \rangle$} \\  
 1 &   3.4310  & 2.0481  & -0.00182 & -2.879 028 767 0519  \\
 2 &   3.5121  & 1.1301  & -0.00225 & -2.879 028 767 3496  \\
 3 &   3.4978  & 1.1267  & -0.00212 & -2.879 028 767 3154  \\
 4 &   3.4988  & 1.0958  & -0.00215  & -2.879 028 767 3196  \\
 5 &   3.5012  & 1.0032  & -0.00215  & -2.879 028 767 31920  \\
    \multicolumn{5}{c}{$\langle \delta \rangle$} \\ 
 1 &   2.4809 &  0.9444   & -0.00535  & 0.155 763 7540   \\ 
 2 &   2.4975 &  0.6872   & -0.00562  & 0.155 763 7156  \\
 3 &   2.5024 &  0.5760   & -0.00560  & 0.155 763 7172 \\
 4 &   2.4989 &  0.6810   & -0.00559  & 0.155 763 7174  \\
    \multicolumn{5}{c}{NO data sequence} \\  
    \multicolumn{5}{c}{$\langle E \rangle$} \\  
 1 &   5.9916  & 2.7712  & -0.2959  &  -2.879 028 767 3054  \\
 2 &   5.9971  & 2.5553  & -0.2995  &  -2.879 028 767 3176  \\
 3 &   5.9959  & 2.6256  & -0.2996  &  -2.879 028 767 3177  \\
 4 &   5.9671  & 4.7513  & -0.2999  &  -2.879 028 767 3179  \\
    \multicolumn{5}{c}{$\langle \delta \rangle$} \\ 
 1 &   3.9973     &  1.620$^*$ & -0.1957  &  0.155 763 7197  \\ 
 2 &   3.9980$^*$ &  1.681$^*$ & -0.1962$^*$  &  0.155 763 7177$^*$  \\
 3 &   3.9997$^*$ &  1.599$^*$ & -0.1963$^*$ &  0.155 763 7175$^*$  \\
    \multicolumn{5}{c}{Energy optimized LTO data sequence} \\  
    \multicolumn{5}{c}{$\langle E \rangle$} \\  
 1 &   5.6562  & 149.13  & -1.543  &  -2.879 028 767 3333  \\
    \multicolumn{5}{c}{$\langle \delta \rangle$} \\ 
 1 &   3.8093$^*$     &  0.3839$^*$ & -0.3879$^*$  &  0.155 763 7191$^*$  \\ 
\end{tabular} 
\end{ruledtabular}
\end{table}
\endgroup

The validity of the series, eq.~(\ref{panalysis}) immediately  
suggests that the asymptotic forms for $\Delta E^N$ are   
\begin{eqnarray} 
\Delta E^N &=& \frac{A_E}{N^{7/2}} + \frac{B_E}{N^{8/2}} + \frac{C_E}{N^{9/2}} + \ldots  
\label{Eseries2} \\ 
\Delta \delta^N &=& \frac{A_{\delta}}{N^{5/2}} + \frac{B_{\delta}}{N^{6/2}} + \frac{C_{\delta}}{N^{7/2}} + \ldots 
\label{dseries} 
\end{eqnarray} 
(In Appendix A it is demonstrated that an exponent variation of 
$p = p_0 + B/\sqrt{N}$ arises from an inverse power series in 
$\Delta X^N$ with a leading term of $B/N^{p_0}$ and with the power increasing 
by $\sqrt{N}$ for successive terms).  Although eqs.~(\ref{Eseries2}) 
and (\ref{dseries}) are best described as a conjecture, the 
numerical evidence in support of the conjecture will be
seen to be strong.  

The applicability and utility of eqs.~(\ref{Eseries2}) and (\ref{dseries})
 was tested by fitting these equations to $\langle E \rangle^N$ and 
$\langle \delta \rangle^{N}$  values and then using eq.~(\ref{remainder}) 
to determine the $N \to \infty$  limits for the individual terms.  Asymptotic 
series with up to 5-terms (i.e. $N_p = 4$) were also investigated.

\begin{figure}[th] 
\centering
\includegraphics[width=8.5cm,angle=0]{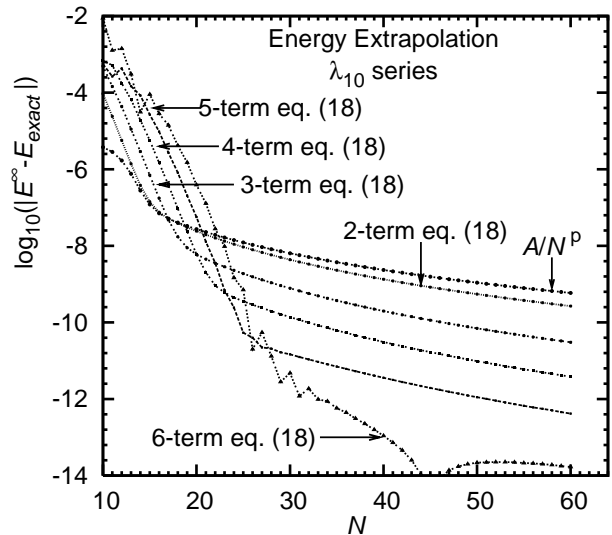}
\vspace{0.1cm}
\caption[]{
The extrapolated $N \rightarrow \infty$ limit for the He ground 
state energy obtained from the different asymptotic expansions.  
The energy sequence from the $\lambda_{10}$ was used. 
The exact $s$-wave energy as given by 
the calculation of Goldman {\em et al} \cite{decleva95a} is
subtracted from the energy.   
}
\label{HeEinf3}
\end{figure}

Figure \ref{HeEinf3} shows $\langle E \rangle^{\infty}$ for the 
$\lambda_{10}$ basis using the asymptotic series of different 
lengths  to estimate the $N \to \infty$ correction.  It is noticeable 
that all the representations of eq.~(\ref{Eseries2}) exhibit 
better convergence properties than eq.~(\ref{Xpdef}) and the 6-term 
representation has the best convergence characteristics for $N>30$.  
(It should be noted that the 3-, 4-, 5- and 6- term extrapolations to 
$\langle E \rangle^{\infty}$ exhibited fluctuations of order 10$^{-5}$ 
to 10$^{-9}$ hartree when the calculation was performed in double
precision arithmetic).  The increasingly better convergence 
characteristics as $N_p$ increases is consistent with  
eq.~(\ref{Eseries2}) being the asymptotic form describing the
energy convergence with respect to a LTO basis.  The estimated
$\langle E \rangle^{\infty}$ limits at $N = 60$ for the various 
asymptotic expressions are given in Table \ref{Nlimits}.  The 6-term
estimate was $\langle E \rangle^{\infty} = -2.879 028 767 31920$ 
hartree which agrees with the value of Goldman, namely 
$E  = -2.879 028 767 31921$ by $1 \times 10^{-14}$ hartree.  The
precision of the extrapolated value exceeds the precision 
of raw $\langle E \rangle^{60}$ energy by a factor of 1,000,000! 
This improvement is best placed in perspective by noting 
that the $\lambda_{10}$ calculation would have to be extended
to $N \approx 10^6$ to achieve the same level of precision.

This extreme accuracy is not reproduced if one uses other 
forms for the asymptotic series.  For example, making the
choice $\Delta E^N = A_E/N^{4} + B_E/N^5 + \ldots$ results
in much poorer estimates of $\langle E \rangle^{\infty}$.  Using
a 4-term series for the $\lambda_{10}$ basis set for this 
asymptotic series gave  
$\langle E \rangle^{\infty} = -2.879 028 802 777$ hartree
which is in error by $3.5 \times 10^{-7}$ hartree.  

The ability to accurately predict $\langle E \rangle^{\infty}$ using 
eq.~(\ref{Eseries2}) has been tested for other values of $\lambda$.  
Making the choice $\lambda = \lambda_{20}$ gave 
$\langle E \rangle^{\infty} = -2.879 028 767 31919$ hartree when the 
6-term series was used to make the extrapolation.  In summary, there 
is strong numerical evidence that eq.~(\ref{Eseries2}) correctly 
describes the convergence of the energy with $N$.   

\begin{figure}
\centering
\includegraphics[width=8.5cm,angle=0]{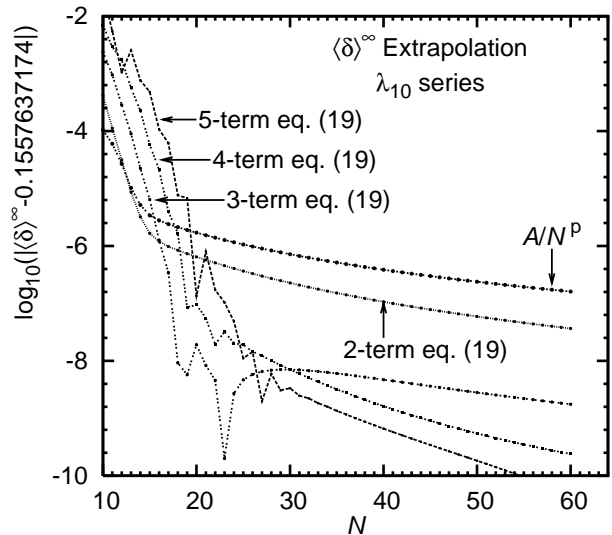}
\vspace{0.1cm}
\caption[]{
The extrapolated $N \rightarrow \infty$ limit for the He ground 
state $\langle \delta \rangle^{\infty}$ for the different 
extrapolation methods applied to the $\lambda_{10}$ basis as 
described in the text.  
}
\label{Hedinf3}
\end{figure}

The slower convergence of the electron-electron $\delta$-function as
$\Delta \delta ^N \approx A/N^{5/2}$ means that the ability to 
extrapolate to the $N \to \infty$ limit is even more important in
obtaining accurate expectation values.    
Figure \ref{Hedinf3} shows the $\langle \delta \rangle^{\infty}$ 
estimates for the $\lambda_{10}$ basis while using eq.~(\ref{Xpdef}) 
and eq.~(\ref{dseries}) to describe the large $N$ limiting behavior.  
It is noticed that the convergence improved as $N_p$ increased 
as long as $N$ was sufficiently large.  Choosing $N > (M+10)$ 
would seem to be sufficient for 2-term or 3-term fits to 
eq.~(\ref{dseries}).  The specific numerical estimates of 
$\langle \delta \rangle^{\infty}$ for various extrapolations at 
$N = 60$ are given in Table \ref{Nlimits}.  The 5-term fit gave 
$\langle \delta \rangle^{\infty} = 0.155 763 7174$ $a_0^3$.  
Given that  $\langle \delta \rangle^{60}$ for the $\lambda_{10}$ 
basis was 0.155 772 7974 $a_0^3$, the improvement in precision 
from the 5-term expansion corresponds to 4-5 orders of magnitude.   
This result is not specific to the $\lambda_{10}$ basis.  Usage 
of the $\lambda_{20}$ basis resulted in an estimate of 
$\langle \delta \rangle^{\infty} = 0.155 763 7175$ $a_0^3$.
Given these results it would seem reasonable to assign a value 
of $0.155 763 7174(2)$ $a_0^3$ to $\langle \delta \rangle$.  
This is the value that was adopted as the ``exact'' value when 
plotting Figures \ref{Hedinf}  and \ref{Hedinf3}.    

The more sophisticated extrapolations shown in Figure \ref{Hedinf3}   
exhibit even-odd fluctuations at the lower values of $N$ and 
led us to omit consideration of a 6-term fit. These
fluctuations are believed to arise from structural features
of the helium ground state for reasons outlined in the next 
section.       
  
\section{Convergence of a Natural Orbital basis}

The asymptotic form of the energy for a wave function written in terms 
of its natural orbital decomposition \cite{shull55a,lowdin56a,carroll79a} 
has also been re-examined for s-wave helium.  First a very large 
calculation of the ground state wave function with a basis of 70 LTOs 
($\lambda = 11.10$) was performed.  The one electron density matrix 
was then diagonalized and the resulting natural orbitals were
used to define a new orbital basis ordered in terms of decreasing  
ground state occupancy.   In its natural orbital form, the
wave function for a $^1S^e$ state is written    
\begin{eqnarray}
|\Psi \rangle = \sum_{i} d_{i} \: \mathcal{A}_{ij} \:
     \langle {\scriptstyle \frac12} \mu_i {\scriptstyle \frac12} \mu_j|0 0 \rangle                                                                                                  
    \phi_i({\bf r}_1) \phi_i({\bf r}_2)  \; .
\label{wvfnno}
\end{eqnarray}
The natural orbital expansion is usually ordered in terms of
decreasing $|d_i|$.   
 
Table \ref{NOtab} gives $\langle E \rangle$ and $\langle \delta \rangle$ 
for the sequence of increasingly larger NO expansions.  For these 
calculations, the generated NOs were added successively and 
$\langle E \rangle^N$ and $\langle \delta \rangle^N$ computed once 
the hamiltonian was diagonalized.  The
calculations were taken up to a maximum NO expansion length of
20.  The LTO basis of dimension 70 was not large enough to give
a precise representation of the NOs beyond that point.  The energies 
in the table are expected to be accurate estimates of the 
``exact'' NO energy for all digits with the possible exception 
of the last two.  The energies in Table \ref{NOtab} are 
slightly lower than the  previous tabulations of the $s$-wave NO 
energies by Carroll {\em et al} \cite{carroll79a} and Goldman 
\cite{goldman95a}.  We treat the NO orbitals merely as a
particularly optimal set of orbitals to input into a CI
calculations.  So unlike Carroll {\em et al} and Goldman 
the configuration space is not restricted to only include
$\phi_i({\bf r}_1) \phi_i({\bf r}_2)$ type configurations.  
It should be noted that we have also done some calculations
using the pure NO configuration space and when this is done
the energies agree with those of Carroll {\em et al} and 
Goldman to all digits.   The $\langle \delta \rangle^N$ 
values in Table \ref{NOtab} are expected to approximate  
those of the ``exact'' basis to about to 10 digits. 

\begin{figure}[th] 
\centering
\includegraphics[width=8.5cm,angle=0]{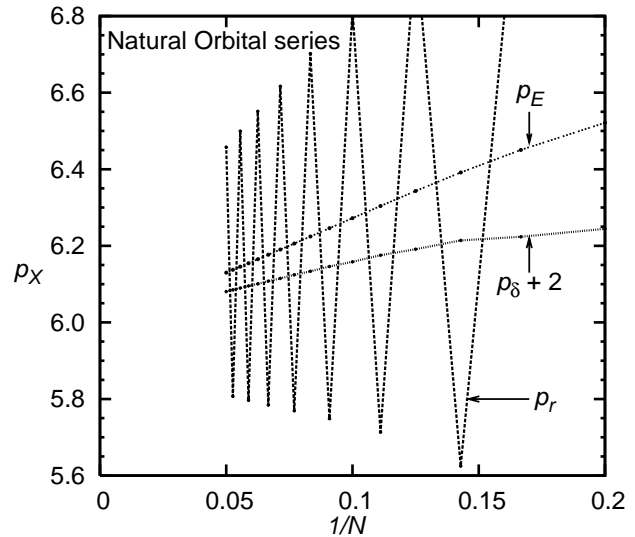}
\vspace{0.1cm}
\caption[]{
The estimated exponents $p_E$, $p_{\delta}$ and $p_r$ as a function of 
$\frac{1}{N}$ for the NO expansion of He ground state. The value
of $p_r$ shows strong even-odd oscillations.   
}
\label{pNOHe}
\end{figure}

Figure \ref{pNOHe} shows the variation of $p_E$ and $p_{\delta}$ 
versus $1/N$ for a sequence of increasingly larger NO calculations 
up to $N = 20$.  The visual inspection of the $p_X$ vs $1/N$ 
curve immediately suggests that $p_E = 6 + A/N + \ldots$ and    
$p_{\delta} = 4 + A/N + \ldots$.  The supposition has been
confirmed by doing fits to the asymptotic form 
\begin{eqnarray} 
p &=& p_0 + \sum_{i=1}^{N_p} \frac{p_i}{N^{i}} \ ,  
\label{panalysis2} 
\end{eqnarray} 
for increasingly larger values of $N_p$.  The results for $p_0$ 
and $p_1$ are given in Table \ref{Nlimits}.  The present calculations 
give $p_0$ values of 5.992, 5.997 and 5.996  (the 5-term series which gave
$p_0 = 5.967$ is likely to be more susceptible to small imperfections 
in the NOs).   A least-squares fit to the function  $p_E = p_0 + p_1/N^t$ 
over the $N \in [11,20]$ interval gave $p_0 = 6.0005$ and $t = 1.070$.  
A least-squares fit to the function  $p_{\delta} = p_0 + p_1/N^t$ 
over the $N \in [11,20]$ interval gave $p_0 = 3.992$ and $t = 0.9174$.  
A small even-odd ripple was present in the $p_{\delta}$ vs $N$ graph.  

The linear variation of $p_X$ vs $1/N$ indicates the asymptotic
series 
\begin{eqnarray} 
\Delta E^N &=& \frac{A_E}{N^{6}} + \frac{B_E}{N^{7}} + \frac{C_E}{N^{8}} + \ldots  
\label{NOEseries} \\ 
\Delta \delta^N &=& \frac{A_{\delta}}{N^{4}} + \frac{B_{\delta}}{N^{5}} + \frac{C_{\delta}}{N^{6}} + \ldots 
\label{NOdseries} 
\end{eqnarray} 
for the variation of the $\Delta E^N$ and $\Delta \delta^N$ with $N$.   
The $O(N^{-6})$ variation of $\Delta E^N$ has been known since the
work of Carroll {\em et al} \cite{carroll79a}.   The present
calculations give a more precise determination of the exponent.  
The value of $p \approx 5.7$ previously reported by 
Goldman \cite{goldman95a} can be discounted.
Besides giving the leading order term with increased 
precision, the present calculations also demonstrate the order of the 
next term in the asymptotic series is $O(N^{-7})$ and thereby strengthen
the justification of asymptotic series based on expansions of the
principal quantum number \cite{klopper99a}.  

The best NO estimate of $\langle E \rangle^{\infty}$ in Table \ref{Nlimits} 
was about an order of magnitude less accurate than that obtained from 
the LTO calculation.   This was not unexpected since the values of 
$\langle E \rangle^N$ in Table \ref{NOtab} are not the ``exact'' NO 
energies, merely very good estimates of these energies.  Also, use of
the NO basis inevitably means a more complicated calculation, and so
it is more likely to be affected by round-off errors and discretization
errors in the numerical quadratures.     

The coefficient of the leading order term for $\Delta E$ was -0.2999 
hartree  which is a bit larger in magnitude than the value initially 
given by Carroll {\em et al} \cite{carroll79a}, namely -0.24 hartree.
This level of agreement is acceptable given the fact that Carroll 
{\em et al} actually use a slightly different $A_E/(N-1/2)^{-6}$ 
functional form (and do not allow for higher order terms) and
extract the value of $A_E$ at $N \approx 10$ which is too 
low to extract the asymptotic value of $A_E$ (the value of 
$A_E$ varies by more than 10$\%$ between $N = 10$ and $N=20$ 
for a single-term asymptotic formula).  The precision of the 
Carroll {\em et al} calculation is also less than that
of the present calculation (they obtained -2.879028765 hartree as 
their variational limit).     

The leading order term for the variation of $\Delta \delta^N$ with $N$
was $O(N^{-4})$.  This dependence is consistent with earlier work of 
Halkier {\em et al} \cite{halkier00a} who found that the variation of 
$\Delta \delta^X$ with $X$ to be $O(X^{-2})$ where $X$ is the principal
quantum number of the natural orbital. When analyzing this set of data
it was discovered that there were regular fluctuations in the derived 
parameters as a function of $N$ as the analysis was made more
sophisticated.  When the 3-term approximation to eq.~(\ref{panalysis}) 
was used the value of $p_0$ oscillated between 3.94 and 4.06 
depending on whether $N$ was even or odd (the oscillations in 
$p_1$ were more marked).  The oscillations became larger for the
4-term fit, here it was found that the $p_0$ typically flipped
between 3.4  and 4.6.  The actual values given in Table \ref{Nlimits} 
were obtained by weighted average, e.g.    
$p_0 = 0.25 p_0(N\!=\!19) + 0.50 p_0(N\!=\!20) + 0.25p_0(N\!=\!21)$.  

These oscillations are most likely due to the physical properties
of the basis set expansion of the He ground state.  It has been known
for a long time that treating the two electrons as an inner and
outer electron can lead to a better description of the radial
correlations \cite{shull56a,sims02a}.   With the electrons having a 
tendency to separate into inner and outer radial orbitals the
possibility does exist that achieving this separation might be
slightly easier or harder if there are an even or odd number
of NOs.  It must be recalled that NOs themselves are objects 
that depend on the electron dynamics. The convergence of the 
mean electron-nucleus distance, i.e.  $\langle r \rangle$ 
was also examined and the convergence pattern  was quite 
irregular.  Defining $p_r$ using eq.~(\ref{pdef}) results in the
strongly oscillating plot of $p_r$ vs $1/N$ observed in Figure 
\ref{pNOHe}.  The oscillations disappear if $\langle r \rangle^N$ 
for only even $N$ (or only odd $N$) are used in a slightly 
modified version of eq.~(\ref{pdef}) and one finds the leading 
order term in $\Delta r^N$ is $O(N^{-6})$.  It should be
noted that similar even-odd fluctuations have also been 
observed in high precision calculations using correlated 
basis sets \cite{schiff65,drake99a,klopper99a}.

\begingroup
\begin{table}[bth]
\caption[]{ The term by term energy and $\langle \delta \rangle$
-function (in $a_0^3$) expectation values for the NO basis.
}
\label{NOtab}
\vspace{0.5cm}
\begin{ruledtabular}
\begin{tabular}{lcc}
$N$ & $\langle E \rangle^N$  & $\langle \delta \rangle^N$   \\ \hline   
  1 &   -2.861 531 101 7265   &    0.190 249 652 529   \\
  2 &   -2.877 929 200 9378   &    0.161 369 548 453   \\
  3 &   -2.878 844 196 5241   &    0.157 747 352 993   \\
  4 &   -2.878 980 288 7909   &    0.156 661 432 897   \\ 
  5 &   -2.879 012 046 8823   &    0.156 240 259 959   \\
  6 &   -2.879 021 844 0177   &    0.156 045 264 861   \\
  7 &   -2.879 025 501 6647   &    0.155 943 428 536   \\                 
  8 &   -2.879 027 069 6581   &    0.155 885 238 797   \\ 
  9 &   -2.879 027 815 8906   &    0.155 849 654 582   \\  
 10 &   -2.879 028 201 2549   &    0.155 826 694 126   \\ 
 11 &   -2.879 028 413 7401   &    0.155 811 228 411   \\      
 12 &   -2.879 028 537 3691   &    0.155 800 434 791   \\               
 13 &   -2.879 028 612 5987   &    0.155 792 675 857   \\               
 14 &   -2.879 028 660 1493   &    0.155 786 956 341   \\
 15 &   -2.879 028 691 2007   &    0.155 782 648 350   \\
 16 &   -2.879 028 712 0592   &    0.155 779 342 132   \\
 17 &   -2.879 028 726 4219   &    0.155 776 762 792   \\
 18 &   -2.879 028 736 5303   &    0.155 774 721 119   \\
 19 &   -2.879 028 743 7842   &    0.155 773 084 071   \\
 20 &   -2.879 028 749 0810   &    0.155 771 756 198   \\
\end{tabular} 
\end{ruledtabular}
\end{table}
\endgroup

The asymptotic behavior of the natural orbital configuration coefficients 
were also determined.  The coefficients are the $d_i$ in eq.~(\ref{wvfnno}).
Assuming that the $d_i$ scale as an inverse power series, 
$d_i \approx A_d/i^p_{\rm NO}$ gives 
\begin{equation}
p_{\rm NO} =   \ln \left(  \frac { d_i^{N}}{d_{i-1}^N} \right) \biggl/   
      \ln \left( \frac{i}{i-1} \right) \ .  
\label{pNOdef} 
\end{equation}
A fit of $p$ to $i$ using the formula   
\begin{eqnarray} 
p_{\rm NO} &=& p_0 + \frac{p_1}{i} + \frac{p_2}{i^2} \ ,   
\label{pNOanalysis} 
\end{eqnarray} 
gave values of $p_0$ that ranged from 3.998 to 4.003 for
successive fits to the 3 previous values for $i$-values 
between 12 and 20 for the $\lambda_{60}$ basis.  It was 
found that    
\begin{eqnarray} 
d_i \approx \frac{0.362}{i^4} + \frac{0.589}{i^5} + \frac{1.492}{i^6} \ ,   
\end{eqnarray} 
at $i = 20$.  Carroll {\em et al} obtained the result   
$d_i \approx \frac{0.271}{(i-1/2)^4}$ \cite{carroll79a}.

\section{Convergence of an Optimized basis}

In this section the convergence properties of the LTO basis which 
is energy optimized at each $N$ are studied.  Developing the 
sequence of exponents $\lambda_M$ that gave the lowest energy 
for a LTO basis of dimension $M$ was tedious.  Defining 
$\delta \langle E \rangle$ and $\delta \langle \delta \rangle$ 
as the differences in $\langle E \rangle$ and 
$\langle \delta \rangle$ arising from an imprecisely known 
$\lambda_M$, one has the relations  
\begin{eqnarray} 
\delta \langle E \rangle & \approx & A (\delta \lambda)^2 \\  
\delta \langle \delta \rangle & \approx & B (\delta \lambda)  \ .   
\end{eqnarray} 
The quadratic dependence of $\delta \langle E \rangle$ with respect 
to  $\delta \lambda$ does make it easier to generate the sequence 
of $\langle E \rangle^N$ values.  But this quadratic dependence upon
$\delta \lambda$ does make it harder to determine $\lambda_M$
since the energy only depends weakly on $\lambda$ in the vicinity 
of the minimum.  Since $\delta \langle \delta \rangle$ depends 
linearly on $\delta \lambda$, any imprecision in $\lambda_M$ 
impacts the  precision of the $\langle \delta \rangle^N$ sequence
more severely.

Some specific data can be used to put this in perspective.  The 
$\lambda_M$ for $M = 1,\ldots,30$ have been determined to a 
precision for $10^{-6}$ for the calculations reported in this 
section.  These gave an energy that was accurate to $10^{-18}$ 
hartree for the $M=15$ calculation, but $\langle \delta \rangle$ 
was only known to a precision of $10^{-11}$ $a_0^3$.  Determination of 
$\langle \delta \rangle$ to a precision of $10^{-15}$ $a_0^3$ would 
require fixing $\lambda_M$ with an accuracy of $10^{-10}$ which 
would necessitate an energy given to a accuracy of $10^{-26}$ hartree.  

\begin{figure}[th] 
\centering
\includegraphics[width=8.5cm,angle=0]{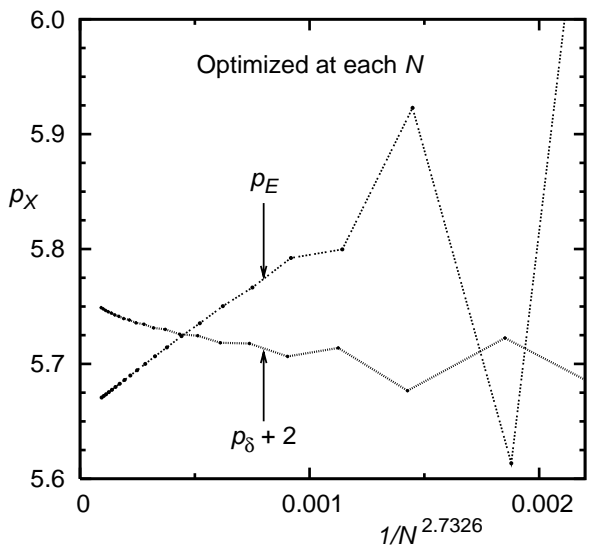}
\vspace{0.1cm}
\caption[]{
The estimated exponents $p_E$ and $p_{\delta}$ as a function of 
$N^{-2.7326}$ for the optimized basis.  The variation of 
$p_{\delta}$ with $N^{-2.7326}$ is not expected to be 
linear.
}
\label{pOPTHe}
\end{figure}

The behavior of $p_E$ and $p_{\delta}$ vs $N$ was sufficiently 
complicated that an initial least squares fit to the equation
$p = p_0 + p_1/N^t$ was performed for $N \in [18,30]$.  The
results of the fit gave 
\begin{eqnarray} 
p_E &=& 5.6562 + \frac{15.69}{N^{2.7326}}  \\   
p_{\delta} &=& 3.8093 - \frac{0.3832}{N^{0.5438}}  \ .    
\label{panalysis3} 
\end{eqnarray} 
The distinctive aspect about the fit is the difference in the leading 
terms of the inverse power series for $p_E$ and $p_{\delta}$.   
Figure \ref{pOPTHe} shows that variation of $p_E$  for the optimized 
LTO basis as a function of $1/N^{2.7326} $ up to $N = 30$.  The
plot of $p_{\delta}$ is tending to curl up for the smallest values
of $1/N^{2.7326}$ because it is not linear in $1/N^{2.7326}$.
 
Another notable feature of Figure \ref{pOPTHe} were the oscillations in 
$p_E$ and $p_{\delta}$ for even and odd values of $N$.  Oscillations
in $p_{\delta}$ were previously seen for the NO sequence but the
$p_{\delta}$ oscillations in Figure \ref{pOPTHe} are more pronounced 
than those in Figure \ref{pNOHe}.  Some of the values in Table 
\ref{Nlimits} were given using the 3-point averaging used previously 
for the Natural Orbital sequence.  

The asymptotic analysis to determine the variational limits were
performed with the following series  
\begin{eqnarray} 
\Delta E^N &=& \frac{A_E}{N^{5.6562}} + \frac{B_E}{N^{8.3888}}   
\label{Eseries3} \\ 
\Delta \delta^N &=& \frac{A_{\delta}}{N^{3.8093}} + \frac{B_{\delta}}{N^{4.3531}} \ . 
\label{dseries2} 
\end{eqnarray} 
The results of the analysis are given in Table \ref{Nlimits}.  
The energy is predicted with an accuracy of $10^{-10}$ hartree
while $\langle \delta \rangle^{\infty}$ is given to an accuracy 
of 10$^{-8}$ $a_0^3$.  Equations (\ref{Eseries3}) and 
(\ref{dseries2}) were not worth extending to include more 
terms.  The power of the next term in eq.~(\ref{Eseries3}) is 
not obvious (refer to the Appendix A) and the oscillations in 
$p_{\delta}$ to a certain extent negate the value of 
extending eq.~(\ref{dseries2}) to include additional terms
(even if we knew what those terms were!). 

\section{Summary and Conclusions}

Results of a sequence of CI calculations of the He ground state 
with an $\ell = 0$ basis have been presented.  This can be regarded
as the simplest model of a real atom that has a correlation cusp.
The energy dependence of the LTO basis was $\Delta E \approx O(N^{-7/2})$.
This rather slow convergence rate can be improved by fitting a 
succession of $\langle E \rangle^N$ values to the inverse power 
series $\Delta E^J = A_E/N^{-7/2} + B_E/N^{-8/2} + \ldots$ and 
estimating the $N \to \infty$ limit.  It ultimately proved 
possible after adopting quadruple precision arithmetic, to 
reproduce the known energy in this model to an accuracy of 
$1 \times 10^{-14}$ hartree.  The specific choice of the 
asymptotic series should be regarded as conjecture supported 
by numerical evidence.  More definite proof would require the 
calculations to be extended to $N > 100$.  The common exponent 
of the LTO basis should not be chosen to optimize the energy 
for the largest calculation since this results in a distorted
convergence pattern.   In effect, optimizing the LTO exponent 
for $N$ LTOs, and then using the $(N\!-\!3)$, $(N\!-\!2)$,
$(N\!-\!1)$ and $N$ energies to determine the coefficients of 
a 3-term expansion to eq.~(\ref{Eseries2}) will give an inaccurate 
estimate of the energy correction needed to achieve the 
variational limit.  Any extrapolation would seem to require 
that $N$ (the number of LTOs) should exceed $M$ (the basis 
dimension at which $\lambda$ was optimized) by about ten or more.  
This conclusion holds for both the energy and electron-electron 
$\delta$-function.  The very slow $O(1/N^{5/2})$ convergence of
$\langle \delta \rangle$ was also circumvented by the use of
the $N \to \infty$ corrections.  

The examinations of the convergence rate for an NO basis set 
revealed a faster convergence.  The NO basis converged
as $O(N^{-6})$ with the next term being $O(N^{-7})$.  The present
determinations of the convergence rates are more rigorous than
those of Carroll {\em et al} \cite{carroll79a}.  One surprising
result was the slight even-odd oscillation in the convergence
of the inter-electronic $\delta$-function.  Examination of the
$\langle r \rangle$ revealed noticeable even-odd oscillations
in $p_{r}$.    The presence of these ripples could complicate 
determination of the variational limit of expectation values other
than the energy.  It was possible to extrapolate the energy of
a 20 orbital NO basis to the variational limit with an accuracy 
of about $10^{-12}$ hartree.     

The convergence rate of the optimized LTO basis was $O(N^{-5.6562})$
with the next term being $O(N^{-8.3888})$.  The degree of uncertainty
in both of these exponents is much larger than for the fixed $\lambda$
LTO sequence or the NO sequence.  The extremely tedious nature 
of the $\lambda$ optimization, combined with the lack of knowledge
about the nature of the asymptotic series beyond the first two 
terms, make this extrapolation a less attractive proposition.    
The noticeable even-odd oscillation in $p_{\delta}$ and even 
$p_E$ further render the method even more unattractive. 
The implications of this behavior are not confined to the present 
work.  For example, it is likely that correlated exponential 
basis sets composed of functions with            
\begin{eqnarray} 
\xi(r_1,r_2,r_{12}) &=& r_1^i r_2^j r_{12}^m \exp(-\lambda r_1) 
\exp(-\lambda r_2) \label{hyllerass} 
\end{eqnarray} 
could also exhibit complicated convergence patterns since $\lambda$ 
is often energy optimized as the basis dimension is increased in size  
\cite{yan05a}.    Consequently, it would not be surprising for  
estimates of the $N \to \infty$ energy correction for variational 
calculations on systems using a Hylleraas basis to be unreliable. 
For example, Yan and coworkers have estimated the variational limit 
in a high precision calculation of PsH using a Hylleraas type basis 
\cite{yan99a}.  Their estimated energy correction for the PsH ground 
state energy (only $9.6 \times 10^{-8}$ hartree) was too small by 
at least a factor of three \cite{mitroy06d}.  

One of the main motivations for the present study was to gain insight
into solving the problems associated with the very slow convergence of 
CI calculations for mixed electron-positron systems 
\cite{bromley02a,bromley02b,mitroy02a,mitroy06a}.  In effect, the 
problem is to determine the complete basis set limit 
\cite{petersson88a,klopper99a,tarczay99a} for these exotic systems.  
The slow $O(N^{-7/2})$ convergence of the energy for an LTO basis set
is greatly improved by the adoption of extrapolation schemes.  Using 
the $N=10$ energy for the $\lambda_{10}$ basis and the best extrapolation 
of the $N=60$ calculation in Table \ref{Nlimits} as two reference points, 
one deduces an effective convergence rate of $O(N^{-10})$.  
The penalty associated with the use of the extrapolation
formulae is the necessity to use quadruple precision arithmetic 
if 3 or more terms are retained in the inverse power series
(note, a 3-term series for $\Delta \delta^N$ was numerically 
stable in double precision arithmetic).  The need to use
the quadruple precision arithmetic is caused by the very
small size of the $\Delta E^N$ increments and the impact of
round-off error on the fit to the inverse power series.       
One somewhat ironic feature is that it is necessary to use
a basis that is {\em not} energy optimized so that the 
extrapolation to the variational limit can be done reliably.

\begin{acknowledgments}

The authors would like to thank Shane Caple and Roy Pidgeon  
of CDU for providing access to extra computing resources,
and Bill Morris of SDSU for computational support.
We would also like to thank David Bosci of Hewlett-Packard (Darwin)  
for giving us access to a demonstration Itanium workstation.      

\end{acknowledgments}

\appendix 
\section{Analysis of the $p$-dependence} 

Let us demonstrate that an asymptotic series  
\begin{equation} 
\Delta X^N = \frac{A}{N^{q}} + \frac{B}{N^{q+t}}  \ldots 
 = \frac{A}{N^{q}}\left(1 + \frac{C}{N^{t}}  \ldots \right)  
\label{XNseries} 
\end{equation} 
(with $C = B/A$) leads to $p = q + F/N^t$ when $p$ is defined
from successive $\Delta X^N$ increments by   
\begin{equation}
p =   \ln \left(  \frac {\Delta X^{N-1}}{\Delta X^N} \right) \biggl/
      \ln \left( \frac{N}{N-1} \right) \ .
\end{equation}  
Substituting $\Delta X^N$ and $\Delta X^{N-1}$ from 
eq.~(\ref{XNseries}) gives   
\begin{equation} 
p  = \ln \left( \frac{ \frac{A}{(N-1)^{q}}\left(1 + \frac{B}{(N-1)^{t}}\right) }
                 { \frac{A}{N^{q}}\left(1 + \frac{B}{N^{t}}\right) } \right)
\biggl/ \ln \left( \frac{N}{N-1} \right) \ . 
\label{dummy} 
\end{equation} 
The logarithm in the numerator can be split into two terms 
\begin{eqnarray} 
\ln \left( \frac{\Delta X^{N-1}}{\Delta X^N} \right) =    
q \ln \left( \frac{N}{N-1} \right) +  
\ln \left(  \frac{ 1\! + \! \frac{C}{(N-1)^{t}}} 
{ 1\! + \! \frac{C}{N^{t}}} \right) 
\label{pdepend} 
\end{eqnarray} 
The first term conveniently cancels with the denominator
to give $q$.  The argument of the second term can be expanded   
\begin{eqnarray} 
 \frac{ 1\! + \! \frac{C}{(N-1)^{t}}} 
{ 1\! + \! \frac{C}{N^{t}}} \! \! & \approx & \! \!    
 \left( 1\! + \! \frac{C}{N^{t}}  + \! \frac{tC}{N^{t+1}} \right)  
 \left(  1\! - \! \frac{C}{N^{t}} \!  + \! \frac{C^2}{N^{2t}} \right)  \nonumber \\    
\! \! & \approx & \! \! 1 + \frac{tC}{N^{t+1}} + \ldots      
\end{eqnarray} 
Using $\ln(1+x) \approx x$ leads to   
\begin{eqnarray} 
\ln \left(  \frac{ 1\! + \! \frac{C}{(N-1)^{t}}} 
{ 1\! + \! \frac{C}{N^{t}}} \right)  & \approx &  \frac{tC}{N^{t+1}}  \ .   
\label{pdepend2} 
\end{eqnarray} 
The denominator is simplified using 
$\ln(N/(N-1)) = \ln(1 + 1/(N-1)) \approx 1/N $ to finally give 
\begin{eqnarray} 
 p = q + \frac{tC}{N^{t}} + \ldots   
\label{pNseries} 
\end{eqnarray} 
as required.  If eq.~(\ref{XNseries}) has successive terms
where the power increments by $t = 1$ or $t = 1/2$ indefinitely,
then this leads to a corresponding series, eq.~(\ref{pNseries}) 
that also have powers that respectively increment by $1$ or $1/2$ 
indefinitely.   This is not necessarily true for arbitrary $t$ in 
eq.~(\ref{XNseries}).

\section{Scaling of the 2-electron integrals} 

The most time-consuming part of the calculation was the generation 
of the electron-electron and annihilation matrix elements.  However,
the expense of this was greatly reduced by generating an initial
set of integrals for a given $\lambda$, and then using a scaling
factor to generate the integral lists for other values of $\lambda$.         

The basic integral that has to be done is 
\begin{eqnarray} 
R(n_a,n_b,n_c,n_d,\lambda)  &=& \iint  dr_1 \ dr_2 \ N_a(\lambda) N_b(\lambda) \nonumber \\ 
    & \times & N_c(\lambda) N_d(\lambda)  f_a(\lambda r_1)  f_b(\lambda r_2) \nonumber \\
    & \times & V(r_1,r_2)   f_c(\lambda r_1)  f_d(\lambda r_2)   
\label{basicR}  
\end{eqnarray} 
All integrals can be defined in terms of $R(n_a,n_b,n_c,n_d,\lambda=1)$. 
Consider the integral (\ref{basicR}) and make the transformation   
$\lambda r = u$.  Therefore $r_1 = u_1/\lambda$ and 
$r_2 =u_2/\lambda$. Similarly $dr_1 = du_1/\lambda$ and 
$dr_2 = du_2/\lambda$ and therefore 
\begin{eqnarray} 
R(n_a,n_b,n_c,n_d,\lambda)  &=& \frac{1}{\lambda^2} \iint  du_1 \ du_2 N_a(\lambda) 
N_b(\lambda) \nonumber \\ 
             & \times & N_c(\lambda) N_d(\lambda)  f_a(u_1)    
 f_b(u_2) \nonumber \\ 
            & \times & V(r_1,r_2)   f_c(u_1)    f_d(u_2)   
\label{basicR2}  
\end{eqnarray} 
From eq.~(\ref{LTOnorm}),  $N_a(\lambda) = \lambda^{1/2} N_a(\lambda=1)$, so   
\begin{eqnarray} 
R(n_a,n_b,n_c,n_d,\lambda) & = &  \iint du_1 \ du_2 N_a(1) N_b(1) \nonumber \\
    &\times &   N_c(1) N_d(1)  f_a(u_1)    f_b(u_2) \nonumber \\ 
    & \times & V(r_1,r_2)   f_c(u_1)    f_d(u_2)   
\label{basicR3}  
\end{eqnarray} 
The scaling for the electron-electron repulsion integral is 
$|{\mathbf r}_1 - {\mathbf r}_2|^{-1} = \lambda |{\mathbf u}_1 - {\mathbf u}_2|^{-1}$.  Hence
\begin{eqnarray} 
R(n_a,n_b,n_c,n_d,\lambda)  =  \lambda R(n_a,n_b,n_c,n_d,1) \ ,      
\label{eescale}  
\end{eqnarray} 
for the electron-electron integral.  When the operator is the $\delta$-function,
one uses the result  
$\delta{(\mathbf r}_1 - {\mathbf r}_2) = \lambda \delta({\mathbf u}_1 - {\mathbf u}_2)$ 
to give 
\begin{eqnarray} 
R(n_a,n_b,n_c,n_d,\lambda)  =  \lambda R(n_a,n_b,n_c,n_d,1)  \ .      
\label{deltascale}  
\end{eqnarray} 
  

\end{document}